\pdfoutput=1
\documentclass[aps,prb,reprint,superscriptaddress,longbibliography]{revtex4-2}

\usepackage{cellspace}
\usepackage{amsmath}
\usepackage{amsfonts}
\usepackage{amssymb}
\usepackage{gensymb}
\usepackage{graphicx}
\usepackage{svg}
\usepackage{bbding}
\usepackage{mathtools}
\usepackage{centernot} 
\usepackage{amsthm} 
\usepackage{cmap}
\usepackage{xcolor}
\usepackage{bigints}
\usepackage{dsfont} 
\usepackage{esint} 
\usepackage{physics}
\usepackage{natbib}
\usepackage{nicefrac} 
\usepackage{cancel}
\usepackage{bm}
\usepackage{upref}
\usepackage{multirow}

\usepackage{varioref} 
\usepackage[]{hyperref} 
\usepackage{url}
\usepackage[capitalise]{cleveref} 

\newcommand{\transp}{\top}

\usepackage{wrapfig}
\usepackage[protrusion=true,tracking=true,expansion,final]{microtype}      
\usepackage{xcolor} 
\usepackage{booktabs} 
\usepackage{subcaption}

\usepackage{chemfig}
\usepackage{adjustbox}

\renewcommand{\cite}[1]{\citep{#1}}

\usepackage[T1]{fontenc}

\vbadness=\maxdimen
\definecolor{citecolor}{HTML}{2980b9}
\definecolor{mydarkblue}{rgb}{0,0.08,0.45}
\definecolor{urlcolor}{rgb}{0,.145,.698}
\definecolor{linkcolor}{rgb}{.71,0.21,0.01}

\hypersetup{ %
	pdftitle={Identifying the topological order of quantized half-filled Landau levels through their daughter states},
	pdfauthor={Evgenii Zheltonozhskii, Ady Stern, Netanel Lindner},
	pdfsubject={},
	pdfkeywords={},
	pdfborder=0 0 0,
	pdfpagemode=UseOutlines,
	bookmarksopen=true,
	breaklinks=true,  
	colorlinks=true,
	urlcolor=urlcolor,
	linkcolor=linkcolor,
	citecolor=citecolor,
	filecolor=mydarkblue,
	pdfview=FitH}

\DeclareFontFamily{OMX}{lmex}{}
\DeclareFontShape{OMX}{lmex}{m}{n}{<-> lmex10}{}

\newcommand{\apsection}[1]{\belowpdfbookmark{#1}{#1} \noindent{\it #1}---\ignorespaces}

\begin{document}


\title{Identifying the topological order of quantized half-filled Landau levels through their daughter states}

\author{Evgenii Zheltonozhskii}
\affiliation{\mbox{Physics Department, Technion, 320003 Haifa, Israel}}
 \email{evgeniizh@campus.technion.ac.il}

\author{Ady Stern}
\affiliation{\mbox{Department of Condensed Matter Physics, Weizmann Institute of Science, Rehovot 7610001, Israel}}

\author{Netanel H. Lindner}
\affiliation{\mbox{Physics Department, Technion, 320003 Haifa, Israel}}


\date{\today}

\begin{abstract}
Fractional quantum Hall states at a half-filled Landau level are believed to carry an integer number $\mathcal{C}$ of chiral Majorana edge modes, reflected in their thermal Hall conductivity.  We show that this number determines the primary series of Abelian fractional quantum Hall states that emerge above and below the half-filling point.  On a particular side of half-filling, each series may originate from two consecutive values of $\mathcal{C}$, but the combination of the series above and below half-filling uniquely identifies $\mathcal{C}$. We analyze these states both by a hierarchy approach and by a composite fermion approach. In the latter, we map electrons near a half-filled Landau level to composite fermions at a weak magnetic field and show that a bosonic integer quantum Hall state is formed by pairs of composite fermions and plays a crucial role in the state's Hall conductivity.

\end{abstract}
\maketitle
The nature of the ground state of quantum Hall states at half-integer fillings, e.g., $\nu=5/2$ \cite{willet1987nu52}, remains an open question of great interest due to the potential for hosting non-Abelian excitations useful for topological quantum computation. Many candidate states \cite{read1999paired,nayak2007nonabelian,hansson2016quantum,ma2022fractional} were proposed, including Moore--Reed Pfaffian \cite{mooreread1991nonabelions}, anti-Pfaffian \cite{levin2007antipfaffian}, particle-hole-Pfaffian (PH-Pfaffian) \cite{son2015composite}, two-component Jain spin-singlet \cite{belkhir1993halfintegral}, and Halperin 331 and 113 \cite{halperin1983theory} states. It was later realized \cite{kane2017pairing} that these states belong to an infinite series of states, all sharing an electrical Hall conductance of $\sigma_{xy}=\frac{e^2}{2h}$ but differing by thermal Hall conductances $\kappa_{xy}=\frac{\pi^2T}{6h}(2+\mathcal{C})$. The topological index $\mathcal{C}$ is an integer; the state is Abelian if $\mathcal{C}$ is even and carries non-Abelian Ising anyons \cite{kitaev2006anyons} if $\mathcal{C}$ is odd. 

While numerical works favor the Pfaffian ($\mathcal{C}=1$) and anti-Pfaffian ($\mathcal{C}=-3$) states \cite{rezayi2009breaking,zaletel2014infinite,rezayi2017landaulevelmixing,rezayi2021energetics} and early quasiparticle tunneling experiments \cite{radu2008quasiparticle} indicated an anti-Pfaffian order, there is now accumulating experimental evidence in favor of the PH-Pfaffian ($\mathcal{C}=-1$) in narrow-well GaAs samples \cite{dolev2011characterizing,banerjee2017observation,dutta2021novel,dutta2021isolated,paul2023topological}, possibly stabilized by disorder \cite{mross2017theory,wang2017topological,simon2020pfaph,hsin2020eftfqh}. Despite numerous proposed and performed experiments \cite{ma2022fractional,li2017denominator,zucker2016phpfaffian,sharma2023compositefermion,spnsltt2019topological,schiller2021extracting,stern2005proposed,bonderson2005detecting,feldman2006detecting,lee2022nonAbelian,yutushui2021identifying,hein2022thermal,manna2022classification,cooper2008observable,yang2009thermopower,haldane2021graviton}, the value of $\mathcal{C}$ and the identification of the half-filled states in different systems are still under intensive study.

A promising direction for this identification is to detect the ``daughter states'': quantum Hall states at nearby filling fractions emerging via anyon condensation mechanism from the state at the half filling (the ``parent state'')
\cite{khveshchenko2006composite,bonderson2007fractional,levin2008collective,hermanns2009condensing}. The Levin--Halperin hierarchies 
\cite{levin2008collective,balram2018anomalfqhe} of states spanning from the Pffafian and anti-Pfaffian states are particularly interesting, as recent experiments in wide GaAs wells \cite{kumar2010nonconventional,singh2023topological} and bilayer graphene  \cite{zibrov2016robust,huang2021valley,assouline2023energy,hu2024studying} observed fractions consistent with the daughter states of the Pfaffian and anti-Pfaffian states arising from these hierarchies. However, it was unknown whether the observed filling fractions were consistent with other parent states as well.

\begin{figure}
    \centering
    \includegraphics[width=\linewidth]{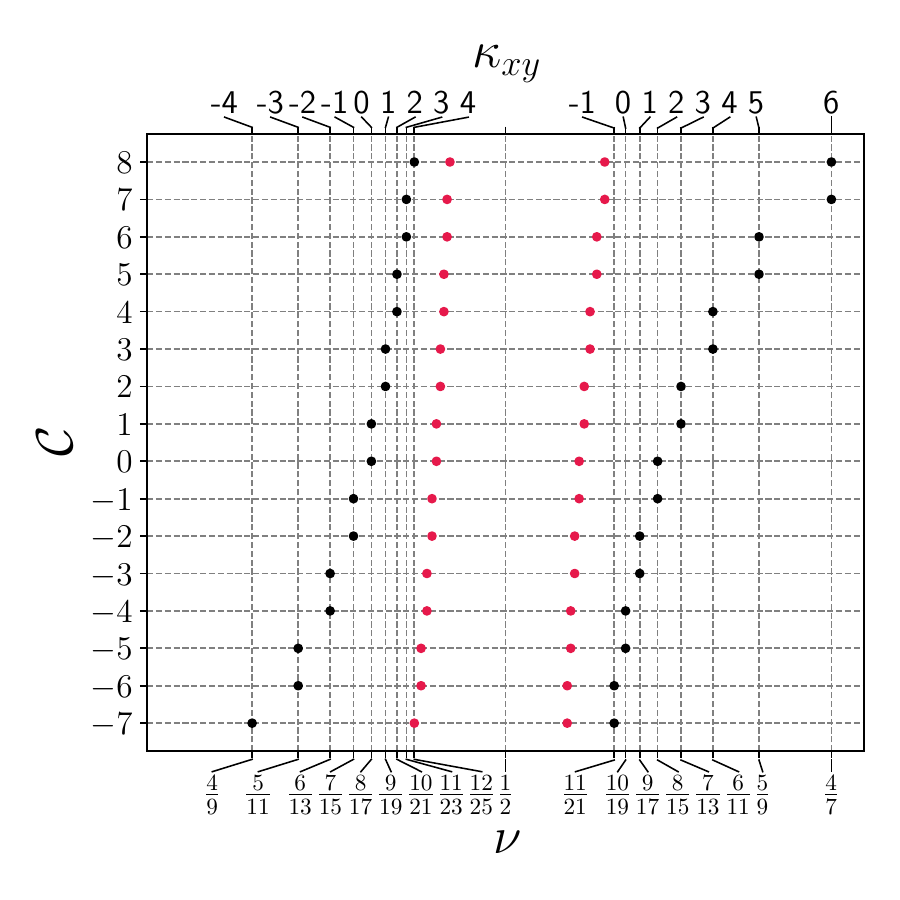}
    \caption{The daughter states at the first two levels of hierarchy. The thermal conductance refers to the state at the first level of the hierarchy (black dots).}
    \label{fig:daughter_states_2}
\end{figure}

In this work, we calculate the series of daughter states that emerge from a parent state with an arbitrary $\mathcal{C}$. While we start the calculation using the hierarchy approach employed in earlier works, we then show how the daughter states can be understood in terms of flux attachment \cite{jain1989cf} that maps electrons at a half-filled Landau level to composite fermions at zero magnetic field. This approach, in which the index $\mathcal{C}$ corresponds to the Chern number of the composite fermion superconductor, allows us to elucidate the relation between the infinite number of possible values of $\mathcal{C}$ and the so-called ``sixteen-fold way'', which introduces a 16-fold periodicity in $\mathcal{C}$ \cite{kitaev2006anyons,ma2019sixteenfold}. 

Specifically, we find that for \textit{Abelian} parent states (even $\mathcal{C}$), the daughter states are parametrized by two integers, $\mathcal{C}/2$ and $m>0$. Denoting $\mathcal{C}' = \mathcal{C}+16k $ for some integer $k$ such that  $-7\leqslant \mathcal{C}' \leqslant 8$, the filling factors satisfy
\begin{align}
\nu^{-1}&=2\pm \qty(8m \pm \nicefrac{\mathcal{C}'}{2})^{-1},
\label{eq:abeliandaughterssummary}
\end{align}
while the thermal Hall conductance is
\begin{align}
\kappa_{xy} &= \qty(\frac{\mathcal{C}}{2}+1\pm 1)\kappa_0, \label{eq:abeliandaughtersthc}
\end{align}
where $\pm$ corresponds to holes and particles and $\kappa_0 = \frac{\pi^2T}{3h}$. 

For non-Abelian parent states (odd $\mathcal{C}$),
each daughter state has the same filling fraction and anyonic content as a daughter state of an Abelian parent state. In particular, for $\nu>1/2$, \cref{eq:abeliandaughterssummary,eq:abeliandaughtersthc} hold for daughter states of the non-Abelian parent states with $\mathcal{C}+1$ instead of $\mathcal{C}$ and
for $\nu<1/2$ with $\mathcal{C}-1$ instead of $\mathcal{C}$.
The filling fractions of the daughter states are visualized in \cref{fig:daughter_states_2}. 

Notably, these results suggest that identifying two series of daughter states---above and below the half-filled Landau level---is sufficient to identify the parent state from which they emerge. An identification of one series is consistent with two values of $\mathcal{C}$, separated by one, corresponding to one Abelian and one non-Abelian state. 
Our results show that the only parent state consistent with the experimental results in wide GaAs wells \cite{singh2023topological} is the Pfaffian state, while for bilayer graphene \cite{zibrov2016robust,huang2021valley,assouline2023energy,hu2024studying}, the only consistent parent state at filling $\nu = n + 1/2$ is the Pfaffian state for odd $n$ and the anti-Pfaffian state for even $n$.

We note that Jain series states may exist at the same filling fractions as the daughter states. However, for the Jain series, we expect to observe the hierarchy of states leading to the filling fraction in question with decreasing energy gaps. Yet, the aforementioned experiments show results inconsistent with these expectations. In fact, a transition between the Jain state regime and the daughter state regime was observed in wide GaAs wells. Additional candidate states exist for some of the observed filling fractions \cite{balram2024fractional}, but they cannot give an alternative explanation for the daughter-state filling fractions appearing both below and above half-filling.

\cref{eq:abeliandaughterssummary} is written in a suggestive way, drawing an analogy to the composite fermion theory of the Jain series~\cite{jain1989cf}. In that theory, $\nu^{-1}=2+\nu_{\text{CF}}^{-1}$, where $2$ is the number of attached flux quanta, $\nu_{\text{CF}}$ is the composite fermions filling factor, and the Jain series corresponds to integer $\nu_{\text{CF}}$. Indeed, all fractions included in (\ref{eq:abeliandaughterssummary}) are Jain fractions. However,  the thermal Hall conductance of a Jain state $\nu=\qty(2+\nicefrac{1}{p})^{-1}$ is $\kappa_{xy}=p\kappa_0$.
Remarkably, in the series we consider here, the variation of the density or magnetic field changes $m$ but keeps $\mathcal{C}$ constant. Thus, it changes the electric Hall conductance while keeping the thermal Hall conductance fixed. As shown below, the $m$-dependence of \cref{eq:abeliandaughterssummary} can be understood as originating from an integer quantum Hall state of bosons comprised of pairs of composite fermions.

The sixteen-fold way is apparent in \cref{eq:abeliandaughterssummary,eq:abeliandaughtersthc}, where shifting $\mathcal{C}$ by 16 yields the same filling, but a thermal Hall conductance that differs by $8\kappa_0$. Since their anyon content is identical, the difference in thermal conductance can be described as originating from the attachment of decoupled layers of the $E_8$ state, a bosonic Abelian state with no anyons \cite{kitaev2011e8,lu2012theory}.

\apsection{Hierarchy construction}
The daughter states of the Abelian parents are obtained from the parent states using the Haldane--Halperin hierarchical construction \cite{haldane1983hierarchy,halperin1984statistics}.

We first review the hierarchical construction.
The wavefunction of the daughter state with electrons at positions $\qty{\vb{r}_k}$ is \cite{hansson2016quantum}
\begin{align}
    \Psi_{\text{d}}\qty(\qty{\vb{r}_k}) &= \int \dd{\vb*{\eta}} \Phi^*\qty(\qty{\vb*{\eta}_j}) \Psi_{\text{p}}\qty(\qty{\vb*{\eta}_j},\qty{\vb{r}_k}),
\end{align}
where $\Psi_{\text{p}}$ is the parent state at filling $1/2$ with $N$ quasiparticles at positions $\qty{\vb*{\eta}_j}$ and $\dd{\vb*{\eta}}= \prod_j \dd{\vb*{\eta}_j}$. 
At a low quasiparticle density, the coordinates $\qty{\vb*{\eta}_j}$ are fixed in a Wigner crystal structure. When the density is sufficiently high, the quasiparticles condense to form the next hierarchy level. Then 
$\Phi$, which is called a pseudo-wavefunction (since it is not single-valued), is
\begin{align}
    \Phi^*\qty(\qty{\vb*{\eta}_j}) &= P\qty(\qty{w_k}) Q\qty(\qty{w_k}) e^{-\sum_k \frac{\abs{q} \abs{w_k}^2}{4\ell_0^2}}. \label{eq:abelian_pwf}
\end{align}
Here, $w = \eta_x \mp i \eta_y$ is a complex coordinate, and the sign depends on the quasiparticle charge sign. The term $Q$ is
\begin{align}
    Q\qty(\qty{w})  = \prod_{j<k} (w_k -w_j)^{\mp 1/\lambda}. 
\end{align}
Under quasiparticle exchange, the wavefunction changes by a phase factor $(-1)^{\pm 1/\lambda}$, as expected from the particles with fractional statistics. $P$ is a symmetric polynomial, which, following the Laughlin argument~\cite{laughlin1983fqh}, is chosen to be
\begin{align}
    P\qty(\qty{w})  = \prod_{j<k} \qty(w_k -w_j)^{2m}
\end{align}
to ensure high-degree zeros when two quasiparticles are brought close together.
$\abs{\Psi_{\text{d}}\qty(\qty{\vb{r}})}^2$ is then describing a two-dimensional plasma at inverse temperature $\beta = \lambda_{\text{d}}$.
The value of $\lambda_{\text{d}}$ is
\begin{align}
\lambda_{\text{d}}&= 2m\pm \lambda^{-1},
\label{eq:lambda_hierarchy}
\end{align}
and from the angular momentum of $\Phi$, $L_{\mathrm{max}}\sim\lambda_{\text{d}} N$, the filling of the quasiparticles is $\nu_{\text{anyon}} = 1/\lambda_{\text{d}}$.
Substituting $q=1/4$, the filling fraction of the state that is formed is 
\begin{align}
    \nu &= \frac{1}{2} \pm \frac{1}{16\lambda_{\text{d}}}. \label{eq:abelian_filling}
\end{align}
We apply this procedure to Abelian $\nu=1/2$ parent states. Using the exchange phase of quasiparticles in the parent state $\lambda^{-1}=\frac{\mathcal{C}+1}{8}$ \cite{kitaev2006anyons}, we get
\begin{align}
\lambda_{\text{d}}&= 2m \mp \frac{\mathcal{C}+1}{8} = \frac{16m \mp (\mathcal{C}+1)}{8} \label{eq:abelian_lambdad} \\
    \nu &= \frac{1}{2}\pm\frac{1}{2} \cdot \frac{1}{16m \mp (\mathcal{C}+1)} =\frac{8m \mp \mathcal{C}/2}{16m \mp (\mathcal{C}+1)},\label{eq:abelian_filling_hierarchy}
\end{align}
where $\mp$ again corresponds to particles and holes.

In the non-Abelian case, the pseudo-wavefunction $\Phi$ depends on the conformal block $\alpha$, which is defined by the pairwise fusion channels:
\begin{align}
   \Psi_{\text{d}}\qty(\qty{\vb{r}}) &= \int \dd{\vb*{\eta}} \sum_\alpha \Phi^*_{\alpha}\qty(\qty{\vb*{\eta}}) \Psi_{\text{p},\alpha}\qty(\qty{\vb*{\eta}},\qty{\vb{r}}). \label{eq:nonab_pseudowf_full}
\end{align}
The pseudo-wavefunction can be split into a product similar to \cref{eq:abelian_pwf} with an additional term, $Y_{\alpha}$, that captures the dependence on $\alpha$:
\begin{align}
    \Phi^*_{\alpha}\qty(\qty{\vb*{\eta}_j}) &= Y_{\alpha}\qty(\qty{w_j})  P\qty(\qty{w_j}) Q\qty(\qty{w_j}) e^{-\sum_j \frac{\abs{q} \abs{w_j}^2}{4\ell_0^2}}.  \label{eq:phialpha}
\end{align}
Similarly to the Abelian case, we want $\Phi^*_{\alpha}$ to transform in opposite manner to the $\Psi_{\alpha}$ under braiding: if $\Psi_{\alpha}\mapsto U_{\alpha \beta}\Psi_{\beta}$, then $\Phi^*_{\alpha} \mapsto \Phi^*_{\alpha} U^*_{\beta\alpha }$. The factor $Y_{\alpha}$ can be expressed as a conformal field theory (CFT) correlator \cite{ginsparg1988applied,difrancesco1997cft}. 

The filling fraction of the daughter state is determined by the maximal angular momentum of $\Phi_{\alpha}$, i.e., by its scaling as $w_i\to \infty$. The scaling of $Y_{\alpha}$ depends on the relative sign of $(\nu-1/2)$ and $\mathcal{C}'$. If the relative sign is negative (i.e., quasiholes for $\mathcal{C}'>0$ and quasiparticles for $\mathcal{C}'<0$), then 
$Y_{\alpha}$ is just a correlator of the Ising theory of the opposite chirality to the one appearing in $\Psi_{\text{p},\alpha}$. For example, for $\mathcal{C}'>0$ the $\alpha$-depending part of $\Psi_{\text{p},\alpha}$ is of the form $\expval{\prod_j \sigma (w_j)}_{\alpha}$. For quasiholes,  $\Phi_{\alpha}$ needs to be anti-holomorphic, thus we can choose $Y_{\alpha} = \expval{\prod_j \sigma' (\bar{w}_j)}_{\alpha}$, which leaves  \cref{eq:nonab_pseudowf_full} invariant under braiding.
Thus, if $(\nu-1/2)$ and $\mathcal{C}'$ have opposite signs, the contribution of the $Y_{\alpha}$ to the angular momentum does not scale with the system size and hence does not affect the filling fraction. The scaling of $\Phi_{\alpha}$ as $w_i\to \infty$ is given by $\lambda_{\text{d}} N$ (\ref{eq:lambda_hierarchy}) with
\begin{align}
    \lambda^{-1} = \frac{\mathcal{C}+1-\mathrm{sgn}(\mathcal{C}') }{8}, \label{eq:lambda_nab_counter}
\end{align}
which is identical to the Abelian parent case with Chern number $\mathcal{C}-\mathrm{sgn}(\mathcal{C}')$.
When $(\nu-1/2)$ and $\mathcal{C}'$ have the same sign, we show below that $Y_{\alpha}$ contributes additional $\mathrm{sgn}(\mathcal{C}') /4$ to $\lambda$, giving a total of 
\begin{align}
    \lambda^{-1} = \frac{\mathcal{C}+1+\mathrm{sgn}(\mathcal{C}') }{8}, \label{eq:lambda_nab_co}
\end{align}
corresponding to the Abelian parent case with Chern number $\mathcal{C}+\mathrm{sgn}(\mathcal{C}')$, from which we get the filling factor using \cref{eq:abelian_filling_hierarchy}.
In the rest of the paper, we limit the values of $\mathcal{C}$ to be between $-7$ and $8$. The discussion can be easily generalized to other values of $\mathcal{C}$.

\apsection{Daughter states of Abelian parent states}
The topological properties of the Abelian daughter states are most concisely described using the matrices $K$ and charge vectors $t$ \cite{wenzee1992structures,wenzee1992classification,wen1995edge,hansson2016quantum}. The $K$-matrices are symmetric and integer-valued, and their determinant counts the topologically distinct quasiparticles. The Hall conductivity is $t^\transp K^{-1}t$. Quasiparticles are described by integer-valued vectors $\ell$; the quasiparticle charge is $t^\transp K^{-1}\ell$, and the mutual fractional statistics of two quasiparticles $\ell_1$, $\ell_2$ is $\ell_1^\transp K^{-1}\ell_2$. The same state can be described by infinitely many pairs of $K$, $t$, related by an $\mathrm{SL}(\mathbb{Z})$ transformation $W$, such that $K'=W^\transp K W$, and $t'=W^\transp t$. 

We use two choices for $K$ and $t$ to describe each state. First, we use a symmetric charge vector $t^\transp=\begin{pmatrix}
    1 & 1 & \dots & 1
\end{pmatrix}$ to construct the $K$-matrices of all Abelian parent states. We start from $113$ state with $K$-matrix 
\begin{align}
    K &= \begin{pmatrix}
        1 & 3\\
        3 & 1
    \end{pmatrix}.
\end{align}
This state has two counterpropagating modes and vanishing thermal Hall conductance, corresponding to $\mathcal{C}=-2$. 

To construct other states, we perform two operations alternately  \cite{ma2019sixteenfold}: particle-hole conjugation  $K \mapsto \begin{pmatrix}
    1 & 0\\
    0 & -K
\end{pmatrix}$ that maps $\mathcal{C}\mapsto -2-\mathcal{C}$, and flipping the neutral modes direction by changing the direction of two fluxes ($K \mapsto \Sigma-K$, where $\Sigma_{ij}=4$) that maps $\mathcal{C}\mapsto -\mathcal{C}$. These operations generate all even-$\mathcal{C}$ states.  

Denoting by $K_\mathcal{C}$ the $K$-matrix of the Abelian parent states with Chern number $\mathcal{C}$, we follow the prescription by \citet{wen1995edge} to construct $K$, $t$ for the daughter states. 
We write the $K$-matrix of the daughter state as
\begin{equation}
   \begin{pmatrix}
        K_\mathcal{C} & \ell \\ \ell^\transp & 2m
    \end{pmatrix}, \label{eq:hier_const}
\end{equation}
where $\ell$ is the vector that generates a quarter-charge quasiparticle. Any choice of $\ell$ with the same charge and statistical phase gives the same state. 

The filling fraction of the state is given by \cref{eq:abelian_filling_hierarchy}. 
The elementary charges are $1/(16m \mp (\mathcal{C}+1))$; the statistical phase is $\pi \pm \frac{2\pi}{16m \mp (\mathcal{C}+1)}$ with $\mp$ and $\pm$ corresponding quasiparticles and quasiholes.
Thermal Hall conductance differs from the parent state by $\pm \kappa_0$, i.e., $\kappa_{xy} = (1+\mathcal{C}/2 \pm 1)\kappa_0$.

The second choice of $K$ and $t$ allows us to describe the daughter states in terms of flux attachment. In this description, the attachment of two flux quanta to each electron maps a daughter state filling fraction to an integer filling fraction. This integer state is composed of a $\abs{\mathcal{C}}/2$ integer quantum Hall (IQH) state of electrons in parallel to a $2m$ IQH state of charge-two bosons \cite{lu2012theory,senthil2012integer,grover2012quantum,lu2012quantum}. To show that, we use the $\mathrm{SL}$ transformations given in supplementary material \cite{SupplementaryMaterial} to change the charge vector to be composed of $\abs{\mathcal{C}}/2$ (2 for $K=8$) entries of $1$ (corresponding to single electrons), and two entries of $2$ (corresponding to bosonic electron pairs), i.e., $t^\transp=\begin{pmatrix}
    1 &\dots & 1 & 2 & 2
\end{pmatrix}$. The $K$-matrix then becomes $K=K_0+\Phi$, where $K_0$ is made of a diagonal fermionic $\frac{\abs{\mathcal{C}}}{2}\times \frac{\abs{\mathcal{C}}}{2}$ block ($2\times 2$ for $K=8$), describing an IQH state of $\nu_f=\mathcal{C}/2$,  and a bosonic $2\times 2$ block 
    \begin{align}
    K_0 &= \begin{pmatrix} K_f & 0\\
    0 & K_b\end{pmatrix} \label{eq:k-flux} \\
    K_b &= \begin{pmatrix}
        0 & 1\\
        1 & 2(1-m)
    \end{pmatrix}.
\end{align}
The matrix $K_b$, together with the corresponding charge vector elements, describe a bosonic IQH state of $\nu_b=2m$, whose contribution to the Hall conductivity is $8m$. 
The flux attachment part of the $K$-matrix is $\Phi=2tt^\transp$, such that $\Phi_{ij}=2t_it_j$. Two flux quanta are attached to each electron, so a boson, which is a pair of electrons, carries four flux quanta. The daughter states of other even-denominator states can be acquired by changing the number of attached flux quanta depending on the denominator, e.g., four per electron at $\nu=1/4$.

As a side note, the parton $\nu=8/17$ state \cite{balram2024fractional}, different from our hierarchical state (it has $\kappa_{xy} = -\kappa_0$ rather than $\kappa_{xy} = 0$),  also belongs to a family of states with the same attached fluxes (\ref{eq:k-flux}). In this case, there are three bosonic modes 
\begin{align}
    K_b &= \begin{pmatrix}
        0 & 1 & 0\\
        1 & 2(1-m) & 0\\
        0&0&2
    \end{pmatrix},
\end{align}
and the charge vector is $t^\transp=\begin{pmatrix}
    1 &\dots & 1 & 2 & 2 & 2
\end{pmatrix}$. For the $\nu=8/17$ state, we have $m=1$ and $\mathcal{C}=-4$; the $\mathrm{SL}$ transformation is given in supplementary material \cite{SupplementaryMaterial}.

For the Jain series, flux attachment in the form of composite fermion theory has been successful in identifying an emergent length scale, the composite fermion cyclotron radius $R_c^*\propto \nu_{\text{CF}}/k_F$, where $k_F$ is the Fermi wavevector \cite{hlr1993halffilled}, that was experimentally observed \cite{willet1993finitewave}. This scale can be written as $\hbar k_F/e^*B$, where $e^*$ is the quasiparticle charge. The mapping we have here suggests the existence of two length scales. The first is inversely proportional to $e^*$, and hence proportional to $16m\mp (\mathcal{C}+1)$. This scale changes as $m$ is varied by changing the distance of the electronic filling fraction from $1/2$. The second scale that emerges from the fermionic $\nu_f=\mathcal{C}/2$ part is $\propto \mathcal{C}/k_F$ and thus is independent of $m$. Interestingly, the corresponding momentum scale, $\hbar k_F/\mathcal{C}$, has a role in the parent state of the half-filled Landau level. It is the momentum scale over which the superconducting order parameter winds for the pairing of angular momentum $\mathcal{C}$. A microscopic model is needed, however, to investigate the roles of these scales, and we leave such an investigation to future work.

\apsection{Daughters of non-Abelian parent states}
The construction of the daughter states for an odd-$\mathcal{C}$ parent state starts from constructing the parent state wavefunction. This can be done using the CFT of the $\mathcal{C}=(2D+1) \mathrm{sgn}(\mathcal{C})$ Ising state  \cite{ma2019sixteenfold}, which includes $D$ neutral bosons $\phi_i$, a downstream charge mode $\phi_\rho$ and an Ising CFT with Majorana mode $\psi$ and spin field $\sigma$. The vertex operators $V_{\beta}(w) = e^{i\beta \phi(w)}$ satisfy
\begin{align}
    \expval{\prod_j V_{\beta_j}(w_j)} &= \prod_{i<j} (w_i-w_j)^{\beta_i \beta_j/k}, \label{eq:vertex_correlator}
\end{align}
and have scaling dimension $\Delta_{\beta} = \beta^2/(2k)$. Here, $\phi_\rho$ has $k=2$, and all other bosonic modes have $k=1$. 

There are $2D+1$ electronic operators
\begin{align}
    \psi_e &= \psi e^{2i\phi_\rho}  \quad 
    \psi_e = e^{\pm i \phi_i} e^{2i\phi_\rho} \label{eq:el}
\end{align}
and $2^D$ quasihole operators
\begin{align}
    \psi_{\text{qh}} &= \sigma  e^{i\phi_\rho/2} \prod_{i=1}^D e^{\pm i \phi_i/2}. \label{eq:qh_operator}
\end{align}
where $\sigma$ is the Ising spin field.
The field $\phi_\rho$ is always holomorphic (downstream), while the direction of $\sigma$, $\psi$, and $\phi_i$ depend on the  $\mathrm{sgn}(\mathcal{C})$. 

The parent wavefunction (\ref{eq:nonab_pseudowf_full}) with electrons at positions $z_k$ and $2N$ excitations at positions $w_j$ fusing to $\alpha$ is then 
\begin{align}
\Psi_{\text{p}, \alpha}\qty(\qty{w_j}, \qty{z_j}) &= 
\Tilde{\Psi}_{\text{p}, \alpha}(w_j,z_k) e^{-\frac{1}{4\ell^2}\sum_k \abs{z_k}^2}  \label{eq:qh_state}\\
\Tilde{\Psi}_{\text{p}, \alpha}(w_j,z_k) &= \expval{\prod_j  \psi_{\text{qh}}(w_j) \prod_k \psi_{\text{e}}(z_k) }_\alpha. \label{eq:qh_correlator}
\end{align}

As mentioned earlier, $\Phi_\alpha$ (\ref{eq:phialpha}) can be expressed as a CFT correlator, specifically the product of a chiral Ising model $\sigma'$ and a chiral boson $\phi'$, in conformal block $\alpha$. For quasiparticle condensate, the fields $\sigma'$ and $\phi'$ are holomorphic and $\Phi_\alpha\qty(\qty{w_j})$ is given by
\begin{align}
    \Phi^*_{\alpha}\qty(\qty{w_j}) &= e^{-\frac{1}{16\ell^2}\sum_j \abs{w_j}^2} \expval{\prod_j \sigma'(w_j) e^{i \sqrt{\lambda_d} \phi'(w_j)}}_\beta R_{ \beta \alpha}, \label{eq:pseudowf_nab}
\end{align}
with the value of $\lambda_d$ from \cref{eq:lambda_hierarchy} and $R_{ \beta \alpha}$ is defined below. For the quasihole condensate, the fields are anti-holomorphic, and $\Phi_\alpha\qty(\qty{\bar{w}_j})$ is obtained from in \cref{eq:pseudowf_nab} with replacing $w_j$ by $\bar{w}_j$ and using $\lambda$ from \cref{eq:lambda_nab_counter}. Consequently, $\sigma$ appearing in \cref{eq:qh_correlator} and $\sigma'$ appearing in $Y_\alpha$ are copropagating if $(\nu-1/2)$ and $\mathcal{C}$ have the same sign, and counterpropagating otherwise.

In the counterpropagating case, $R_{\alpha \beta}=\delta_{\alpha \beta}$; since $\sigma$ and $\sigma'$ have opposite chirality, they also have opposite braiding phases, which sum up to zero. 
In the copropagating case,  $\sigma$ and $\sigma'$ have the same chirality, and the phases do not sum up to zero (correlator involving $\sigma'$ transforms with $U$ rather than $U^\dagger$). 
To fix that, we apply $R_{\alpha \beta}$ given by the unique unitary matrix such that $R^{-1} U R = U^\dagger$. In the conformal block $\alpha$, the $\sigma$ particles are grouped into pairs, and every pair fuses to $1$ or $\psi$. The matrix $R$ swaps between these fusion channels ($1\leftrightarrow \psi$). The total phase is then always $-\frac{\pi}{8}+\frac{3\pi}{8}=\frac{\pi}{4}$. This is the source of the $\mathrm{sgn}(\mathcal{C}') /4$ factor in \cref{eq:lambda_nab_co}.
The filling fractions are then obtained using \cref{eq:lambda_hierarchy,eq:abelian_filling}.

The full expression for the daughter state wavefunction is obtained by substituting \cref{eq:qh_state,eq:pseudowf_nab} into \cref{eq:nonab_pseudowf_full}.
The sum over $\alpha$ results in the full correlator without branch cuts:
\begin{align}
\Psi_{\text{d}}\qty(\qty{z_k}) &= \int \dd{w} \Tilde{\Psi}_{\text{d}} \cdot e^{-\frac{1}{4\ell^2}\sum_i \abs{z_k}^2 - \frac{1}{16\ell^2}\sum_i \abs{w_j}^2} 
\label{eq:full_daughter_correlator}\\
\Tilde{\Psi}_{\text{d}}\qty(\qty{z_k})&=\expval{\prod_j  \tilde{\psi}_{\text{qh}}(w_j) \prod_k \tilde{\psi}_{\text{e}}(z_k) }, \label{eq:full_correlator}
\end{align}
where $\tilde{\psi}_{\text{qh}}$ and $\tilde{\psi}_{\text{e}}$ are the modified operators acquired by gathering all terms depending on quasihole and electron coordinates, correspondingly, and $\dd{w} = \prod_j \dd{w_j}$.

The thermal Hall conductance is given by the central charge of the total CFT in \cref{eq:full_correlator}. For the parent state, $\kappa_{xy} = \qty(1+\mathcal{C}/2)\kappa_0$. The thermal Hall conductance of the added CFT (consisting of chiral boson $\phi'$ and Ising CFT $\sigma'$) is $\pm 3\kappa_0/2$; in total, $\kappa_{xy} = (5+\mathcal{C})\kappa_0/2 $ for quasiparticles and $\kappa_{xy} = (\mathcal{C}-1)\kappa_0/2$ for quasiholes.

Daughter state excitations correspond to the insertion of an operator $\psi_{\text{ex}}$ into \cref{eq:full_correlator}. For the correlator to be single-valued and non-singular,
 $\psi_{\text{ex}}$ should be local relative to the modified quasihole $\tilde{\psi}_{\text{qh}}$ and electron $\tilde{\psi}_e$ operators (given below). 
 By writing the operator-product expansion  $\psi_{\text{ex}}(v)\tilde{\psi}_{\text{qh}}(w) \sim (v-w)^{\Delta_{\text{fused}} - \Delta_{\text{ex}}- \Delta_{\text{qh}}}\psi_{\text{fused}}$, we conclude that the scaling dimension difference $ \Delta_{\text{fused}} - \Delta_{\text{ex}}- \Delta_{\text{qh}}$ should be a non-negative integer for the wavefunction to be single-valued and non-singular. A similar procedure is applied to $\psi_{\text{ex}}(v)\tilde{\psi}_{\text{e}}(w)$.

In the counter-propagating case,  $\tilde{\psi}_{\text{qh}}$ is given by
\begin{align}
    \tilde{\psi}_{\text{qh}}(w_j,\bar{w}_j) &= \sigma(w_j, \bar{w}_j) e^{i\phi_\rho/2 +i\sqrt{\lambda_d}  \phi'}
     \prod_i e^{\pm i\phi_i/2} \label{eq:counter_mod_qh}
\end{align}
where $\sigma(w_j, \bar{w}_j)$ is the spin field in the nonchiral Ising model, and $\tilde{\psi}_{\text{e}}=\psi_{\text{e}}$.

We write a general operator in the combined CFT, $\psi_{\text{ex}} = \chi e^{i (a\phi_\rho + \sum_j a_j \phi_j + b\phi')}$ for $\chi \in \qty{1,\psi, \sigma, \mu}$, where $\mu$ is the disorder operator of Ising CFT.
Denoting $b=(b'-a/4-\sum_j a_j/2)/\sqrt{\lambda_d}$, the excitations local relative to electrons and quasiholes have the following constraints: $a$ and $a_j$ are integer if $\chi \in \qty{1, \psi}$ and half-integer otherwise, $b'$ is integer if $\chi \in \qty{1, \sigma}$ and half-integer otherwise. 

When satisfying the above constraints, the excitation $\psi_{\text{ex}}$ can be written as a product of $\mathcal{O} = \psi e^{i\phi'/2\sqrt{\lambda_d}}$ and additional $\tilde{\psi}_{\text{qh}}$ and $\tilde{\psi}_e$. The elementary charge can be determined \cite{levin2008collective} by inserting $\mathcal{O}^2$ into \cref{eq:full_daughter_correlator}; using \cref{eq:vertex_correlator} and $\psi^2=1$, this results in factor $\prod_j (w_0-w_j)$. That means that  $\mathcal{O}^2$ creates a Laughlin quasihole in the condensate. Since the condensed anyon charge  is $e/4$, $\mathcal{O}^2$ creates the charge of $\nu_{\text{anyon}} e/4$, and using $\nu_{\text{anyon}} = \frac{1}{2m + (2D\pm 1)/8 }$, we find the elementary charge to be $\frac{e}{16m + 2D\pm 1 }$. Calculating $\expval{\mathcal{O}(w_1)\mathcal{O}(w_2)}$ using \cref{eq:vertex_correlator} and $\expval{\psi(w_1)\psi(w_2)} = (w_1-w_2)^{-1}$ we find that topological phase is $\pi \pm \frac{2\pi}{16m + 2D\pm 1 }$.

In the copropagating case, we use an identity relating correlators of Ising CFT and of a chiral bosonic field  $\phi''$:
\begin{align}
   &\expval{\prod_j \sigma'(w_j)}_\alpha  \expval{\prod_j \sigma(w_j) \prod_k \psi(z_k)}_\beta R_{\alpha \beta} = \nonumber\\& \quad\quad \quad =\expval{\prod_j\exp(i\phi''(w_j)/2)\prod_k\cos(\phi''(z_k))}. \label{eq:alphabeta}
\end{align}
This identity can be understood in two steps. First, the bosonization of the Dirac fermion formed by the two Majorana fermions $\psi$, $\psi'$ \cite{tam2019coupled} gives $\psi \sim \cos(\phi'')$. Second, the fusion channel of the two particles $(\sigma(w_1), \sigma'(w_1))$ and $(\sigma(w_2), \sigma'(w_2))$  is either $\qty(\psi,1)$ or $\qty(1,\psi')$, giving after summation the full Dirac fermion  $\exp(i\phi'')$, which can be written as $\exp(i\phi''(w_1)/2)\exp(i\phi''(w_2)/2)$. Thus, we get
\begin{align}
 \tilde{\psi}_{\text{qh}}(w_j,\bar{w}_j) &= e^{\frac{i}{2}\phi_\rho+i\sqrt{\lambda_d} \phi'+i\phi''/2}
     \prod_i e^{\pm i\phi_i/2} \label{eq:coprop_qh}\\
     \tilde{\psi}_{\text{e}}(z_k) &= e^{2i\phi_\rho(z_k)} \cos(\phi''(z_k)). \label{eq:coprop_el}
\end{align}
In \cref{eq:coprop_qh}, $\phi_\rho$ is holomorphic, and $\phi'$, $\phi''$, $\phi_i$ are holomorphic for $\mathcal{C}>0$ and anti-holomorphic otherwise.

Repeating the procedure outlined above, we get the elementary excitation $\mathcal{O} = e^{i\qty(\phi'/2\sqrt{\lambda_d}-\phi'')}$, with the elementary charge of $\frac{e}{16m - 2D-2\pm 1 }$ and topological phase $\pi \pm \frac{2\pi}{16m - 2D-2\pm 1 }$.

Since all the topological properties of the daughters of the non-Abelian states are identical to those of the daughters of the corresponding Abelian states ($\mathcal{C}+1$ for $\nu>1/2$ and $\mathcal{C}-1$ for $\nu<1/2$), we conclude that these are indeed the same states.

To summarize, we constructed the daughter states of quantized paired states of half-filled Landau levels. We showed that the daughter states formed around half-filling reflect the Chern number of the neutral modes of the half-filled state from which they emerge. Provided that no unexpected phase transition occurs as the filling is varied away from the half-filled level, the daughter states can be used to identify the half-filled state topological order.

\begin{acknowledgments}
We thank David Mross, Misha Yutushui, and Yuval Ronen for instructive discussions. EZ is supported by the Adams Fellowships Program of the Israel Academy of Sciences and Humanities. NL is grateful for funding from the ISF Quantum Science and Technology program (2074/19). AS was supported by grants from the ERC under the European Union’s Horizon 2020 research and innovation programme (Grant Agreements LEGOTOP No. 788715), the DFG (CRC/Transregio 183, EI 519/71), and by the ISF Quantum Science and Technology (2074/19). 
\end{acknowledgments}

\bibliography{main}


\onecolumngrid

\appendix{}

\renewcommand\thefigure{A.\arabic{figure}} 
\renewcommand\thetable{A.\arabic{table}} 
\renewcommand\theequation{A.\arabic{equation}}


\setcounter{figure}{0}  
\setcounter{table}{0}
\setcounter{equation}{0}

\section{$W$ matrices between hierarchical and flux basis}
\paragraph{Hierarchical states}
To transform the $K$ matrix of the Abelian state of the daughter state to the flux basis, we first write them in hierarchical basis by transforming $K_{\mathcal{C}}$ with $W=\delta_{i,j}-\delta_{i+1,j}$ and applying procedure from \cref{eq:hier_const}.
Now we give here explicitly the transition matrices $W$ for every dimension of the $K$ matrix that map the hierarchical basis to the flux-attached basis, i.e., to $K_0$ defined in 
\cref{eq:k-flux}.
\begin{align}
    W_3 &= \begin{pmatrix}
        1&2&2\\
        0&1&1\\
        0&0&1
    \end{pmatrix}\\
    W_4 &= \begin{pmatrix}
        1&1&2&2\\
        0&-1&-2&-2\\
        0&0&-1&-1\\
        0&0&0&1
    \end{pmatrix}\\
    W_5 &= \begin{pmatrix}
        1&1&1&2&2\\
        0&-1&-1&-2&-2\\
        0&0&1&2&2\\
        0&0&0&1&1\\
        0&0&0&0&-1
    \end{pmatrix}\\
    W_6 &= \begin{pmatrix}
        1&1&1&1&2&2\\
        0&-1&-1&-1&-2&-2\\
        0&0&1&1&2&2\\
        0&0&0&-1&-2&-2\\
        0&0&0&0&-1&-1\\
        0&0&0&0&0&-1
    \end{pmatrix}
\end{align}

\paragraph{Parton state}
The transition matrix that maps between the $K$-matrix given in Eq.~(5) in Ref.~\cite{balram2024fractional} and $K$-matrix in flux basis (\ref{eq:k-flux}) is given by
\begin{align}
    W_p = 
\begin{pmatrix}
 1 & 1 & 2 & 2 & 2 \\
 0 & 0 & 1 & 0 & 1 \\
 0 & 0 & 0 & 1 & -1 \\
 0 & 0 & 0 & 0 & 1 \\
 0 & 1 & 1 & 1 & 1 \\
\end{pmatrix}
\end{align}

\section{Experimental data}
\cref{tab:fillings} summarizes the experimental observation of states at fillings corresponding to daughter states.

\begin{table}
    \centering
    \caption{Experimental data on observed states at fillings corresponding to daughter states}
    \begin{tabular}{lcccc} \toprule
        \textbf{Reference} & \textbf{Material} & \textbf{Filling fraction} & \textbf{Observed fractions} & \textbf{Candidate} \\
        &  & & \textbf{near half-filling} & \textbf{parent state} \\\midrule
         \citet{kumar2010nonconventional} & GaAs & 5/2 &6/13 & Inconclusive\\
         \citet{singh2023topological} & GaAs & 1/2 & 8/17 and 7/13& Pfaffian\\
         \citet{zibrov2016robust} & Graphene& -1/2 & 8/17 and 7/13 & Pfaffian\\
         \citet{huang2021valley} & Graphene& 3/2& 8/17 and 7/13 & Pfaffian\\
         \citet{huang2021valley} & Graphene& 5/2 & 6/13 and 9/17 & Anti-Pfaffian\\
         \citet{assouline2023energy} & Graphene& -1/2 & 8/17 and 7/13 & Pfaffian\\
         \citet{hu2024studying} & Graphene& -1/2& 8/17 and 7/13 & Pfaffian\\
         \citet{hu2024studying} & Graphene& 3/2& 8/17 and 7/13 & Pfaffian\\
         \citet{hu2024studying} & Graphene& 7/2 & 8/17 and 7/13& Pfaffian\\
         \bottomrule
    \end{tabular}
    \label{tab:fillings}
\end{table}

\section{Excitations of daughter state CFTs}

In this section, we give a detailed derivation of the elementary excitations of the daughter states of the non-Abelian parent states. We tackle the counterpropagating ($(\nu-1/2)$ and $\mathcal{C}'$ have different signs) and copropagating cases ($(\nu-1/2)$ and $\mathcal{C}'$ have the same sign).

The process consists of two steps: first, we find all possible excitations of the resulting CFT  $\psi_{\text{ex}}$ by requiring
 $\psi_{\text{ex}}$ to be local relative to the modified quasihole $\tilde{\psi}_{\text{qh}}$ and electron $\tilde{\psi}_e$ operators. This condition is equivalent to the requirement that the scaling dimension difference $ \Delta_{\text{fused}} - \Delta_{\text{ex}}- \Delta_{\text{qh}}$ (in case of quasiholes) is a non-negative integer.

 Second, we determine an operator $\mathcal{O}$ such that any $\psi_{\text{ex}}$ can be written as a product of $\mathcal{O}$ and additional $\tilde{\psi}_{\text{qh}}$ and $\tilde{\psi}_e$.

\subsection{Counterpropagating case}
\paragraph{Possible excitations}
We use the fact that the scaling dimensions of a fermion is $1/2$ and of a vertex operator $e^{i\beta \phi}$ is $\beta^2/(2k)$, where $k=2$ for $\phi_\rho$ and $k=1$ for the rest of the bosonic fields $\phi'$ and $\phi_i$, and that the scaling dimension of $\sigma$ is $1/8$. Using the above, we obtain the scaling dimensions of the electron, quasihole, and $\psi_{\text{ex}}$
\begin{align}
    \Delta_{\text{e}} &= 1/2 + 1 = 3/2 \\\Delta_{\text{qh}} &= 1/8+1/16+\lambda_d/2 + D/8 \\
    \Delta_{\text{ex}} &= \Delta_{\chi} + a^2/4 + \sum_j a_j^2/2 + b^2/2,
\end{align}
We now write down possible fusion between a $\psi_{\text{ex}}$ and the two different electron operators, $\psi_{\text{e}1}$ and $\psi_{\text{e}2}$, as well as quasihole $\tilde{\psi}_{\text{qh}}$, where we denote fusion result between particles $\psi_a$ and $\psi_b$ as $\psi_{a,b}$. These are given by
\begin{align}
    \psi_{\text{ex}, \text{e}1} &= \chi e^{i (a\phi_\rho + \sum_j a_j \phi_j + b\phi')} \psi e^{2i\phi_\rho} = \chi \psi  e^{i ((a+2)\phi_\rho + \sum_j a_j \phi_j + b\phi')} \\
    \psi_{\text{ex}, \text{e}2} &= \chi e^{i (a \phi_\rho + \sum_j a_j \phi_j + b\phi')} e^{\pm \phi_k }e^{2i\phi_\rho}= \chi e^{i ((a+2) \phi_\rho + \sum_j (a_j \pm \delta_{jk}) \phi_j + b\phi')} \\
    \psi_{\text{ex}, \text{qh}} &= \chi e^{i (a\phi_\rho + \sum_j a_j \phi_j + b\phi')} \sigma e^{i\phi_\rho/2 +i\sqrt{\lambda_d} \phi'}
     \prod_k e^{\pm i\phi_k/2} = \chi \sigma e^{i ((a+1/2)\phi_\rho + \sum_j (a_j\pm 1/2) \phi_j + (b+\sqrt{\lambda_d})\phi')} 
\end{align}
Note that only one specific quasihole operator condenses, and thus, without loss of generality, we can choose the signs in front of $a_j$ in the quasihole operator to be positive.
We now calculate the scaling dimensions:
\begin{align}
    \Delta_{\text{ex}, \text{e}1} &= \Delta_ {\chi \psi}  + (a+2)^2/4 + \sum_j a_j^2/2 + b^2/2 \\
    \Delta_{\text{ex}, \text{e}2} &=  \Delta_{\chi}+(a+2)^2/4  + \sum_j (a_j \pm \delta_{jk})^2/2 + b^2/2 \\\Delta_{\text{ex}, \text{qh}} &= \Delta_{\chi\sigma }+\qty(a+1/2)^2/4 + \sum_j (a_j+ 1)^2/2 + (b+\sqrt{\lambda_d})^2/2
\end{align}
Subtracting the scaling dimensions of the fused particle and its constitutions, we get
\begin{align}
   \Delta_{\text{ex}, \text{e}1}-\Delta_{\text{ex}} - \Delta_{\text{e}} &= \Delta_ {\chi \psi}  + (a+2)^2/4 + \sum_j a_j^2/2 + b^2/2 - \qty[\Delta_{\chi} + a^2/4 + \sum_j a_j^2/2 + b^2/2 + 3/2] = \nonumber \\&= \qty(\Delta_ {\chi \psi} -\Delta_{\chi} ) + a - 1/2\\ 
   \Delta_{\text{ex}, \text{e}2}-\Delta_{\text{ex}} - \Delta_{\text{e}} &= \Delta_{\chi}+(a+2)^2  + \sum_j (a_j \pm \delta_{jk})^2/2 + b^2/2  - \qty[\Delta_{\chi} + a^2/4 + \sum_j a_j^2/2 + b^2/2 + 3/2] = \nonumber  \\&= 
   a+1  +  (1 \pm 2a_k)/2 - 3/2 = a \pm a_k\\
   \Delta_{\text{ex}, \text{qh}}-\Delta_{\text{ex}} - \Delta_{\text{qh}} &= \Delta_{\chi\sigma }+\qty(a+1/2)^2/4 + \sum_j (a_j+ 1/2)^2/2 + (b+\sqrt{\lambda_d})^2/2 - \nonumber \\&- \qty[\Delta_{\chi} + a^2/4 + \sum_j a_j^2/2 + b^2/2 + 1/8+1/16+\lambda_d/2 + D/8] = \nonumber \\&=
   \qty(\Delta_{\chi\sigma } - \Delta_{\chi} ) + a/4 + 1/16 + \sum_j (1/4+ a_j)/2 + (2b\sqrt{\lambda_d}+ \lambda_d)/2 -3/16 -\lambda_d/2-\abs{C}/16 = \nonumber \\&=
      \qty(\Delta_{\chi\sigma } - \Delta_{\chi} ) + a/4 + \sum_j  a_j/2 + b\sqrt{\lambda_d} -1/8
\end{align}
Substituting the definition $b = (b'-a/4- \sum_j  a_j/2)/\sqrt{\lambda_d}$, we get 
\begin{align}
   \Delta_{\text{ex}, \text{e}1}-\Delta_{\text{ex}} - \Delta_{\text{e}} &= \qty(\Delta_ {\chi \psi} -\Delta_{\chi} ) + a - 1/2 \label{eq:fusionexe1counter}\\ 
   \Delta_{\text{ex}, \text{e}2}-\Delta_{\text{ex}} - \Delta_{\text{e}} &= 
   a+1  +  (1 \pm 2a_k)/2 - 3/2 = a \pm a_k  \label{eq:fusionexe2counter}\\
   \Delta_{\text{ex}, \text{qh}}-\Delta_{\text{ex}} - \Delta_{\text{qh}} &= 
      \qty(\Delta_{\chi\sigma } - \Delta_{\chi} )  + b' -1/8 \label{eq:fusionexqhcounter}
\end{align}
Using the fusions of the Ising part \cite{difrancesco1997cft}
\begin{align}
    \Delta_{\chi \psi} -\Delta_\chi -\frac{1}{2}&= \begin{cases}
        0 & \chi = 1\\
        -1 & \chi=\psi\\
        -\frac{1}{2} & \chi \in \qty{\sigma, \mu}
    \end{cases}
\end{align}
we get from the requirement that RHS of the \cref{eq:fusionexe1counter,eq:fusionexe2counter} is integer that $a$ and $a_j$ are integer if $\chi \in \qty{1,\psi}$ and half-integer if  $\chi \in \qty{\sigma,\mu}$.
The same requirement on \cref{eq:fusionexqhcounter} along with
\begin{align}
    \Delta_{\chi\sigma} -\Delta_{\chi } -\frac{1}{8} &= \begin{cases}
        0 & \chi \in \qty{1,\sigma}\\
        -1/2 & \chi \in \qty{\psi,\mu}
    \end{cases}
\end{align}
means that $b'$ is integer if $\chi \in \qty{1,\sigma}$ and half-integer if $\chi \in \qty{\psi,\mu}$. In total, the excitation can be written as
\begin{align}
    \psi_{\text{ex}} = \chi \exp(i \qty(a\phi_\rho + \sum_j a_j \phi_j + \qty(b'-a/4- \sum_j  a_j/2)\phi'/\sqrt{\lambda_d} )) \label{eq:excounter}
\end{align}
with $a$, $a_j$, $b'$ integer or half-integer as determined above.  
\paragraph{Elementary excitations}
We now ask which elementary excitation can generate any operator given by \cref{eq:excounter}. We claim that this can be done using the operator $\mathcal{O} = \psi \exp(i\phi'/2\sqrt{\lambda_d})$. Indeed,  consider an operator of the form $ \mathcal{O}^{n_o} \tilde{\psi}_{\text{e}}^{n_{\text{e}}} \tilde{\psi}_{\text{qh}}^{n_{\text{qh}}}$. Equating it to $\psi_{\text{ex}}$ (ignoring for a moment $a_j$ which can be matched by choosing the appropriate quasihole operator), we get a set of linear equations:
\begin{align}
a&= 2n_{\text{e}} + n_{\text{qh}}/2\\
b&= \sqrt{\lambda_d} n_{\text{qh}} + n_o/(2\sqrt{\lambda_d})
\end{align}
which has a solution 
\begin{align}
    n_{\text{qh}}  &= 2(a-2n_{\text{e}})\\
    n_o &= 2b' -a(16m+k+1)/2 + (16m+k)n_{\text{e}},
\end{align}
where $k$ is $\mathcal{C}+1+\mathrm{sgn}(\mathcal{C}')$ and thus is odd, and $n_e$/$n_{\text{qh}}$ matches the required value of $\chi$.

\subsection{Copropagating case}
In the copropagating case, the excitations are given by \cref{eq:coprop_qh,eq:coprop_el}.
Writing the generic excitation as 
\begin{align}
   \psi_{\text{ex}} &= \exp(i\qty(a\phi_\rho+b\phi' + c\phi'' + \sum_j a_j \phi_j)),
\end{align}
we repeat the procedure of determining local excitations. Scaling dimensions $\Delta_{\text{e}}$ and     $\Delta_{\text{qh}}$ are the same as in previous case:
\begin{align}
    \Delta_{\text{e}} &= 3/2 \\
    \Delta_{\text{qh}} &= 1/8+1/16+\lambda_d/2 + D/8 \\\Delta_{\text{ex}} &=  a^2/4 + \sum_j a_j^2/2 + b^2/2 +c^2/2
\end{align}
The scaling dimensions of fusions are
\begin{align}
     \Delta_{\text{ex}, \text{e}1} &= (a+2)^2/4 + (c+1)^2/2 + b^2/2+ \sum_j a_j^2/2 \\
     \Delta_{\text{ex}, \text{e}1} &= (a+2)^2/4 + c^2/2 + b^2/2+ \sum_j (a_j\pm 1)^2/2 \\
     \Delta_{\text{ex}, \text{qh}} &= (a+1/2)^2/4 + (c+1/2)^2/2 + (b+\sqrt{\lambda_d})^2/2+ \sum_j (a_j+1/2)^2/2.
\end{align}
After the subtraction, we get
\begin{align}
   \Delta_{\text{ex}, \text{e}}-\Delta_{\text{ex}} - \Delta_{\text{e}} &=  a + c \label{eq:fusionexe1co}\\ 
   \Delta_{\text{ex}, \text{qh}}-\Delta_{\text{ex}} - \Delta_{\text{qh}} &= 
     b'  + c \label{eq:fusionexqhco}
\end{align}

We note that these are exactly the same expressions as in the \cref{eq:fusionexe1counter,eq:fusionexqhcounter} in the previous section, up to the absence of the Ising field part and the replacement of the fermion $\psi$ by the vertex operator of $\phi''$ (coefficient $c$). As such, the same solution applies here (substituting $e^{-i\phi''}$ instead of $\psi$), and the elementary excitation is $\mathcal{O} = e^{i\qty(\phi'/2\sqrt{\lambda_d}-\phi'')}$.

\end{document}